\documentclass[twocolumn,showpacs,preprintnumbers]{revtex4}
\usepackage{amssymb}
\usepackage{amsfonts}
\usepackage{amsmath}
\usepackage{graphicx}
\usepackage{dcolumn}
\usepackage{bm}

\input{tcilatex}

\begin{document}

\title{Quantum theory as a relevant framework for the statement of
probabilistic and many-valued logic}
\author{E. D. Vol}
\email{vol@ilt.kharkov.ua}
\affiliation{B. Verkin Institute for Low Temperature Physics and Engineering of the
National Academy of Sciences of Ukraine 47, Lenin Ave., Kharkov 61103,
Ukraine.}
\date{\today }

\begin{abstract}
Based on ideas of quantum theory of open systems we propose the consistent
approach to the formulation of logic of plausible propositions. To this end
we associate with every plausible proposition diagonal matrix of its
likelihood and examine it as density matrix of relevant quantum system. We
are showing that all logical connectives between plausible propositions can
be represented as special positive valued transformations of these matrices.
We demonstrate also the above transformations can be realized in relevant
composite quantum systems by quantum engineering methods. The approach
proposed allows one not only to reproduce and generalize results of
well-known logical systems (Boolean, Lukasiewicz and so on) but also to
classify and analyze from unified point of view various actual problems in
psychophysics and social sciences.
\end{abstract}

\maketitle

\section{Introduction}

In the last decades the rapidly growing field of quantum information and
quantum computation aimed at the creation of quantum computer won high
priority in quantum physics \cite{1a}. It is well known that all possible
applications of the theory (quantum cryptography, quantum
teleportation,quantum seach algorithms and so on)are based on the
fundamental concepts of superposition and entanglement of quantum states
.Unfortunately quantum coherence and entanglement are the properties that
very sensitive to the interaction of quantum system with its environment.
Precisely this reason makes the creation of quantum computer such
extraordinary difficult problem. On the other hand the concept of density
matrix and formalism of quantum theory of open systems (QTOS) allows one
from unified point of view to describe classical ( that is ordinary
statistical) correlations as well as quantum ones. In particular as was
shown by author earlier \cite{2b}, provided that certain restrictions on the
form of interaction of quantum system in question with its environment are
imposed, classical correlations may be entirely selected from quantum ones
and therefore studied independently.In present paper we exploit this idea by
somewhat unexpected way for formulation of logic of plausible propositions
that is probabilistic logic(PL). Apart from methodological, new approach has
undoubted heuristic advantages, because it allows one to apply unified
framework to describe the behavior of various classical systems with
discrete variables. We hope that approach proposed will find application in
psychophysics for example for the better understanding of color vision laws
and especially in social sciences where probabilistic judgements and complex
undetermined behavior are ubiquitous both for individuals and groups. We are
going to devote in future a few publications to these problems but in this
paper we examine the only illustrative example namely 1) algebraic model of
human reflexion choice by V.A. Lefebvre \cite{3b} and propose its possible
generalization . The rest of the paper is organized as follows. In Sect.1 we
remind some necessary facts from the Boolean calculus of propositions and
introduce the main concept of our approach, namely , diagonal matrix of
likelihood (LM) for any plausible proposition (PP). This matrix can be
interpreted as density matrix of relevant quantum system. After that we are
showing that under such interpretation all logical connectives between PP
can be represented as special positive valued transformations of diagonal LM
conserving both their traces and diagonal form .In Sect.2 we apply our
approach as a proper tool to formulate the concepts and operations of
many-valued probabilistic logic (MVPL). In particular we study more detail
the case of three-valued or Lukasiewicz PL in view of its greater simplicity
and importance. It may be noted that our approach allows one also to operate
at the same time with propositions whose LM have different dimensions. This
property of PL removes the usual logic borders and essentially expends its
possibilities for applications. In Sect.3 we demonstrate that all operations
of PL can be physically realized with the help of operators acting on the
states of subsystems of the relevant quantum system provided that we
appropriately specify the form of its interaction with environment. This
reason allows one to implement a set of parallel logical computations in
subsystems of composite quantum system without worryng about destructive
action of decoherence on this process Of course we have in mind only
classical computations in quantum system. In Sect.4 in the capacity of
illustration we consider the only example of PL application. Namely we
examine\ from the PL point of view\ \ concrete psychological theory: the
Lefebvre model of human reflexion choice \cite{3b} and propose its possible
generalization .

Now let us go to the presentation of our results.

\section{Quantum approach to the basic concepts of probabilistic logic}

In the beginning let us recall some necessary facts from the Boolean
calculus of logical propositions. In the Boolean theory is assumed that
every proposition "$A$" may be either true or false and hence one can
introduce a discrete variable $x_{A\text{\ }}$ which takes only two values:
1, if "$A$" is true and 0 in the opposite case. Starting with some primary
collection of propositions one can construct step by step with the help of
logical connectives more and more complex propositions. There are three
basic connectives : "not", "and", "or" which can be defined or with the help
of truth tables or, that is more preferred, by Boolean functions in the
following way. Let we have single proposition "$A$" with Boolean variable $%
x_{A}$ then proposition "not $A$" or "$\overline{A}$ " has the Boolean
function $x_{\overline{A}}=1-x_{A\text{ }}$, and if we have two propositions
"$A$" and "$B$" with variables $x_{A}$ and $x_{B}$ respectively, then
Boolean function for proposition "$A$ and $B$" is defined as \ $x(A$ and $B$%
) = $x_{A}x_{B}$ , and for proposition "$A$ or $B$ " $\ $as $x(A$ or $B$ ) = 
$x_{A}+x_{B}-x_{A}x_{B\text{ }}$. Below we will need certain helpful result
from the Boolean calculus which allows one by inductive way to obtain
arbitrary Boolean function of any number of variables. This result reads as
follows: any Boolean function of $N+1$ variables $-h\left(
x_{1},...x_{N+1}\right) $ has the only representation in the form $%
h(x_{1},...x_{N+1})=x_{N+1}f(x_{1,}....x_{N})+\overline{x_{N+1}}%
g(x_{1},...x_{N})$, \ where $f$ \ and $g$ \ some Boolean functions of $N$
variables and $\overline{x}\equiv 1-x$. This result implies that tne number
of Boolean functions of N variables- $A_{N}$\ \ satisfies \ to the relation
: $A_{N+1}=A_{N}^{2}$. \ Taking into account that exist only two functions 0
and 1of zero variables we result in that $A_{N}=2^{2^{N}}$. It is clear
however that both in daily life and in social sciences we meet at every step
with the propositions which are only more or less plausible and hence cannot
be described by the Boolean calculus. Precisely for such cases we propose to
take concepts of QTOS as a framework to establish general rules for
operating with PP and to formulate the laws of PL. To this end we associate
with every PP "$A$" some $2\times 2$ diagonal matrix:\ $\rho \left( A\right)
=%
\begin{pmatrix}
p &  \\ 
& 1-p%
\end{pmatrix}%
$- the likelihood matrix (LM) of "$A$". Value of $p$ can be considered
naturally as probability that proposition "$A$" is true. The advantage of
such representation consists in possibility to consider LM as density matrix
(DM)\ of relevant quantum system and hence to apply powerful machinary of
QTOS for determination of probabilistic logic required laws. The first step
in this direction should be the determination of forms for LM responding to
basic logic connectives of PL. It is clear that LM of \ PP "not $A$" should
be in the form

$\rho (\overline{A\text{ })}=%
\begin{pmatrix}
1-p &  \\ 
& p%
\end{pmatrix}%
$. This matrix can be represented as $\rho \left( \overline{A}\right) =U\rho
\left( A\right) U^{T}$ , where $U$ is the matrix $%
\begin{pmatrix}
0 & 1 \\ 
1 & 0%
\end{pmatrix}%
$ and notation $U^{T}$ means a matrix transposed to $U$. Remind here the
helpful result from QTOS namely the Kraus representation which claims that
density matrix of open quantum system at any time $\rho \left( t\right) $
can be written as:%
\begin{equation}
\rho \left( t\right) =\sum\limits_{i=1}^{N}K_{i}\left( t\right) \rho \left(
0\right) K_{i}^{+}\left( t\right) ,  \label{v1}
\end{equation}%
where a collection of operators- $\left\{ K_{i}\left( t\right) \right\} $
should satisfy to the restriction $\sum\limits_{i=1}^{N}K_{i}^{+}\left(
t\right) K_{i}\left( t\right) =1$ that provides the normalization of $\rho
\left( t\right) $. We will need for our purpose analog of the Kraus
representation but \ with certain modifications because in our case the
required class of admissible transformations must conserve not only traces
of LM but their diagonal form as well. Omitting minor technical details let
us formulate the final decisive result : Let $\rho $ is $N\times N$ diagonal
matrix with positive elements whose trace is equal to 1 and let $G$ is some $%
M\times N$ matrix which possesses two following properties 1) all elements
of $G$ are $0$ or $1$ and 2)in each column of $G$ the only element is equal
to 1 and the rest are equal to zero.In this case the admissible
transformation $\widetilde{\rho }=G\rho G^{T}$ maps $\rho $ on $M\times M$
diagonal matrix $\widetilde{\rho }$ with positive diagonal elements that
satisfy to the relation -$\sum\limits_{i=1}^{M}\widetilde{\rho _{ii}}=1$.
The proof of this result can be obtained by direct verification. Armed with
this basic result we can find the required form of LM for arbitrary number
of plausible propositions and for any connectives between them. It should be
emphasized that for this purpose we need to take tensor product of all
propositional LM as the initial ("incoherent") state for further operations.
This incoherent state obviously diagonal in our case. Then by applying of
certain admissible transformation we can obtain LM for required PP. To
demonstrate how this rule works in practice we construct basic connectives
between two PP. Let us begin with the PP "$A$ and $B$". Let \ \ $\rho \left(
A\right) =%
\begin{pmatrix}
p &  \\ 
& 1-p%
\end{pmatrix}%
$ is LM of the PP "$A$" and $\rho \left( B\right) =%
\begin{pmatrix}
q &  \\ 
& 1-q%
\end{pmatrix}%
$ is LM of the PP "$B$" respectively and let us define as $G_{and}$ the $%
2\times 4$ admissible matrix-$%
\begin{pmatrix}
1 & 0 & 0 & 0 \\ 
0 & 1 & 1 & 1%
\end{pmatrix}%
$. Acting in accordance with the above rules we obtain%
\begin{eqnarray}
\rho \left( A\text{ }and\text{ }B\right) &=&G_{and}\rho \left( A\right)
\otimes \rho \left( B\right) G_{and}^{T}=  \label{v2} \\
&=&%
\begin{pmatrix}
pq &  \\ 
& 1-pq%
\end{pmatrix}%
.  \notag
\end{eqnarray}%
Similarly if we take as transformation matrix $G_{or}=%
\begin{pmatrix}
1 & 1 & 1 & 0 \\ 
0 & 0 & 0 & 1%
\end{pmatrix}%
$ we result in that LM of PP "$A$ or $B$" is%
\begin{eqnarray}
\rho \left( A\text{ }or\text{ }B\right) &=&G_{or}\rho (A)\otimes \rho \left(
B\right) G_{or}^{T}=  \label{v3} \\
&=&%
\begin{pmatrix}
p+q-pq &  \\ 
& \left( 1-p\right) \left( 1-q\right)%
\end{pmatrix}%
.  \notag
\end{eqnarray}%
It is easy to check directly that in the case when $p$ and $q$ take values 0
and 1 only, accepted LM matrices are entirely consistent with ordinary
Boolean functions for connectives "and", "or" (provided that we identify LM -%
$%
\begin{pmatrix}
1 & 0 \\ 
0 & 0%
\end{pmatrix}%
$ with true and LM-$%
\begin{pmatrix}
0 & 0 \\ 
0 & 1%
\end{pmatrix}%
$ with false propositions respectivly). In addition note that in the case of
two propositions all 16 admissible matrices G with above-mentioned
properties exactly correspond to 16 Boolean functions of two variables .
This correspondence between probabilistic logical connectives (that is a
collection of all admissible transformations $\left\{ G_{ad}\right\} $ ) and
all Boolean functions remains true in the case of any number of variables.
In particular the number of admissible transformations for N probabilistic
propositions as one can easily verify directly coincides with the number of
Boolean functions of N propositions that is equal to $2^{2^{N}}$.

\section{Many-valued probabilistic logic from quantum point of view}

In this part \ we are going to generalize concepts and constructions of PL
on the case of many-valued probabilistic logic .The major attention will be
drawn to the case of three-valued PL in view of its greatest simplicity and
importance.However all results obtained in this part can be easily
generalized on the case of arbitrary MVPL. So, in the case of three-valued
PL as well as in ordinary three-valued or Lukasiewicz logic \cite{4b} we
suppose that all propositions can be divided on three disjoint classes:
true-(T), undetermined-(U) and false-(F),but unlike "deterministic"
Lukasiewicz logic now we believe that for every proposition "$A$" we know
only probabilities :$p_{1}\equiv p_{T},$ $p_{2}\equiv p_{U},$ $p_{3}\equiv
p_{F}$ for proposition "$A$" to be a member of a given class. As before we
associate with every 3 valued PP its LM which in present case takes the
form: $\rho \left( A\right) =diag\left( p_{1},p_{2},p_{3}\right) $ provided
that $p_{1}+p_{2}+p_{3}=1$. Now we construct three basic connectives " not",
"and", "or" for 3 valued PP. To this end we will again exploit the class of
admissible transformations with required properties according to the above
basic rule: $\widetilde{\rho }=G_{ad}\rho G_{ad}^{T}$. For example
connective "not" will be defined as $\rho \left( \overline{A}\right)
=G_{not}\rho \left( A\right) G_{not}^{T}=diag\left( p_{3},p_{2},p_{1}\right) 
$ with $G_{not}=%
\begin{pmatrix}
0 & 0 & 1 \\ 
0 & 1 & 0 \\ 
1 & 0 & 0%
\end{pmatrix}%
$. Connective "$A$ and $B$" for two PP- "$A$" \ with LM $\ \rho \left(
A\right) =diag\left( p_{1},p_{2},p_{3}\right) $ and "$B$" with LM- $\rho
\left( B\right) =diag\left( q_{1},q_{2},q_{3}\right) $ could be defined as%
\begin{widetext}
\begin{equation}
\rho \left( A\text{ and }B\right) =%
\begin{pmatrix}
p_{1}q_{1} &  &  \\ 
& p_{1}q_{2}+p_{2}q_{1}+p_{2}q_{2} &  \\ 
&  & p_{3}+q_{3}-p_{3}q_{3}%
\end{pmatrix}.%
\label{v4}
\end{equation}
\end{widetext} which can be represented as

$\rho \left( A\text{ and }B\right) =G_{and}\cdot \rho \left( A\right)
\otimes \rho \left( B\right) \cdot G_{and}^{T}$ with $G_{and}=$ $=%
\begin{pmatrix}
1 & 0 & 0 & 0 & 0 & 0 & 0 & 0 & 0 \\ 
0 & 1 & 0 & 1 & 1 & 0 & 0 & 0 & 0 \\ 
0 & 0 & 1 & 0 & 0 & 1 & 1 & 1 & 1%
\end{pmatrix}%
$. In addition PP - "$A$ or $B$" may be defined as%
\begin{widetext}
\begin{equation}
\rho \left( A\text{ or }B\right) =%
\begin{pmatrix}
p_{1}+q_{1}-p_{1}q_{1} &  &  \\ 
& p_{2}q_{2}+p_{2}q_{3}+p_{3}q_{2} &  \\ 
&  & p_{3}q_{3}%
\end{pmatrix}.%
\label{v5}
\end{equation}
\end{widetext}with $G_{or}=%
\begin{pmatrix}
1 & 1 & 1 & 1 & 0 & 0 & 1 & 0 & 0 \\ 
0 & 0 & 0 & 0 & 1 & 1 & 0 & 1 & 0 \\ 
0 & 0 & 0 & 0 & 0 & 0 & 0 & 0 & 1%
\end{pmatrix}%
$. We does not claim that the above definitions are the only possible in
three valued PL (moreover it is not the case). But they have essential
advantages in view of at least three reasons:1) let all probabilities $%
p_{i},q_{i}(i=1,2,3)$ take only two values 0 and 1, then accepted
definitions consistent with rules of ordinary Lukasiewicz calculus \cite{4b}
if one identify the PP-$diag\left( 1,0,0\right) $ with true, the PP -$%
diag\left( 0,1,0\right) $ with undetermined and the PP-$diag\left(
0,0,1\right) $ with false propositions respectively 2) It is easy to see
that both connectives 'and", "or" are symmetric that is "$A$ and $B$"$=$"$B$
and $A$" and" $A$ or $B$"$=$"$B$ or $A$". 3) In addition two important De
Morgan's duality laws are valid for the accepted definitions as well as in
the Boolean calculus, namely:a) Not ("$A$" and $B$")= (Not "$A$") or (Not" $%
B $") and \ b) Not("$A$ or $B$") = (Not "$A$") and (Not "$B$"). In this
connection it is worth noting that some fundamental laws of ordinary logic
as for example "The Law of Excluded Middle" do not valid in probabilistic
logic. It is clear also that such logical connective as for example
implication can be obtained from the basic ones according to the usual
relation:"$A\Longrightarrow B$" = "$\overline{A}$ or $B$" that implies: 
\begin{widetext}
\begin{equation}
\rho \left( A\Longrightarrow B\right) =%
\begin{pmatrix}
p_{3}+q_{1}-p_{3}q_{1} &  &  \\ 
& p_{2}q_{2}+p_{2}q_{3}+p_{1}q_{2} &  \\ 
&  & p_{1}q_{3}%
\end{pmatrix}%
. 
\label{v6}
\end{equation}
\end{widetext}with $G_{imp}=%
\begin{pmatrix}
1 & 0 & 0 & 1 & 0 & 0 & 1 & 1 & 1 \\ 
0 & 1 & 0 & 0 & 1 & 1 & 0 & 0 & 0 \\ 
0 & 0 & 1 & 0 & 0 & 0 & 0 & 0 & 0%
\end{pmatrix}%
.$We would like to emphasize that using the admissible transformations one
can also construct various logical operations with propositions of different
types. It is possible for example to connect two valued PP with LM - $\rho
\left( A\right) =%
\begin{pmatrix}
p &  \\ 
& 1-p%
\end{pmatrix}%
$ and three valued PP $\ $\ with LM-$\rho \left( B\right) =diag\left(
p_{1},p_{2},p_{3}\right) $. In our opinion such possibility is not purely of
academic interest but essentially expends the borders of applying PL to
concrete problems. One important example we can specify : it is the
phenomenon of color vision in humans.It is well known that perception of
color in humans is performed in the eye by three types of cones.Although
approximately 90 percents of people are trichromats that is they perceive
all three base colors (red, green and blue),nevertheless about 10 percents
(males as a rule) are dichromats that is one type of their cones does not
operate properly.It is clear that in general there are three types of
dichromats and it is really the case.We believe that method of admissible
transformations of the present paper can help one to establish the explicit
connection between trichromat and dichromat color perceptions and we are
going to develop this line in postponed publication.

\section{Quantum engineering of the probabilistic logic constructions}

\bigskip In previous parts we introduce basic concepts and operations of PP
calculus by somewhat abstract way with the help of certain class of
admissible transformations.Here we want to demonstrate how these formal
constructions can be realized in relevant quantum composite system by the
quantum engineering tools.To get acquainted with modern quantum enginering
technology and its possibilities we recommend \cite{5b} and references
therein.Our consideration will be based on well-known Lindblad equation that
describes an evolution of open Markov quantum system. Let $\rho \left(
t\right) $ is a density matrix of open system in question then the Lindblad
equation for evolution of $\rho \left( t\right) $ has the following general
form \cite{6b}%
\begin{equation}
\frac{d\rho }{dt}=-\frac{i}{\hbar }\left[ H,\rho \right] +\sum\limits_{i}\ %
\left[ R_{i}\rho ,R_{i}^{+}\right] +h.c.,  \label{v7}
\end{equation}%
where $H$ is a Hermitian operator ("hamiltonian" of open system,) and $R_{i}$
are collection of non-Hermitian operators that describe its interaction with
environment. Below we need the next helpful result that was proven by author
earlier \cite{2b}. Let us assume that $\ H=0$ and all operators $R_{i}$ \
have the monomial form, namely: $R_{i}=k_{i}f_{\alpha _{i}}^{+}f_{\alpha
_{2}....}^{+}f_{\beta _{1}}f_{\beta _{2}}....$ ( we assume here that quantum
system in question is described by the algebra of Fermi operators $\
f_{i}^{+},f_{j}$) \ Then the equations of evolution for the diagonal
elements of DM , namely- $\rho \left( N_{1,}N_{2}...\right) $ (where $%
N_{i}=f_{i}^{+}f_{i}$ are occupation numbers ) can be entirely separated
from nondiagonal ones and hence studied independently. We will exploit this
result to demonstrate how probabilistic logic concepts can be realized by
quantum engineering tools. Our first step in this direction is constructing
the required initial state that is tensor product of two LM which correspond
to PP "$A$" and PP "$B$" respectivly. Let $\ \rho \left( A\right) =%
\begin{pmatrix}
p &  \\ 
& 1-p%
\end{pmatrix}%
$and\ $\rho \left( B\right) =%
\begin{pmatrix}
q &  \\ 
& 1-q%
\end{pmatrix}%
$are LM of propositions in question. Let us consider two-qubit open quantum
system which evolution is governed by the Lindblad equation with the
collection of four operators: $R_{1}=\sqrt{\frac{a}{2}}f_{1}^{+}$, $R_{2}=%
\sqrt{\frac{b}{2}}f_{2}^{+}$, $R_{3}=\sqrt{\frac{c}{2}}f_{1}$, $R_{4}=\sqrt{%
\frac{d}{2}}f_{2}$. Then Eq. (\ref{v2}) for the diagonal elements of DM of
the system of interest takes the form:%
\begin{widetext}
\begin{equation}
\frac{d\rho _{N_{1}N_{2}}}{dt}=a\left( N_{1}\rho _{\overline{N_{1}}N_{2}}-%
\overline{N_{1}}\rho _{N_{1}N_{2}}\right) +b\left( N_{2}\rho _{N_{1}%
\overline{N_{2}}}-\overline{N_{2}}\rho _{N_{1}N_{2}}\right) +c\left( 
\overline{N_{1}}\rho _{\overline{N_{1}}N_{2}}-N_{1}\rho _{N_{1}N_{2}}\right)
+d\left( \overline{N_{2}}\rho _{N_{1}\overline{N_{2}}}-N_{2}\rho
_{N_{1}N_{2}}\right).
\label{v8}
\end{equation}
\end{widetext}where the notion $\overline{N_{i}}=1-N_{i}$ is used. After the
simple algebra we can find the stationary solution of Eq. (\ref{v8}) which
reads as:%
\begin{widetext}
\begin{equation}
\rho _{00}=\frac{cd}{\left( a+c\right) \left( b+d\right) },\rho _{01}=\frac{%
bc}{\left( a+c\right) \left( b+d\right) },\rho _{10}=\frac{ad}{\left(
a+c\right) \left( b+d\right) },\rho _{11}=\frac{ab}{\left( a+c\right) \left(
b+d\right) }
\label{v9}
\end{equation}
\end{widetext}It is easy to see that for arbitrary values of a, b, c, d the
stationary DM of \ composite system in question is the tensor product of
their subsystems DM, namely: $\rho _{st}=\rho _{1}\otimes \rho _{2\text{ \ }%
} $(where $\rho _{1}=\frac{1}{a+c}%
\begin{pmatrix}
a &  \\ 
& c%
\end{pmatrix}%
$ and $\rho _{2}=\frac{1}{b+d}%
\begin{pmatrix}
b &  \\ 
& d%
\end{pmatrix}%
$)$.$Thus if one let $\frac{a}{a+c}=p$ and $\frac{b}{b+d}=q$ we come to the
result that stationary DM of composite system coincides with tensor product
of two propositional LM. So after such preparation procedure we have in
hands required initial state for further quantum manupulations. Let us
demonstrate for example how can be realized the state of the two- qubit
system in which the state of first subsystem corresponds to LM of PP- "$A$
or $B$", and the state of second subsystem corresponds to LM of PP \ "$A$
and $B$". The needed manipulation for this purpose is as follows. To this
end consider again Eq. (\ref{v2}) for two-qubit open system with single
operator :$R=\sqrt{\frac{a}{2}}f_{2}^{+}f_{1}$.In this case the Lindblad Eq.
(\ref{v2}) takes the form%
\begin{equation}
\frac{d\rho }{dt}=a\left( N_{2}\overline{N_{1}}\rho _{\overline{N_{1}}\text{ 
}\overline{N_{2}}}-\overline{N_{2}}N_{1}\rho _{N_{1}N_{2}}\right) \text{.}
\label{v10}
\end{equation}%
It is easy to see that stationary solution of Eq. (\ref{v10}) can be
represented in the form $\rho _{00}\left( t\right) =\rho _{00}\left(
0\right) =\left( 1-p\right) \left( 1-q\right) ,\rho _{11}\left( t\right)
=\rho _{11}\left( 0\right) =pq.$In addition Eq. (\ref{v10}) implies that $%
\rho _{10}$ tends to zero and $\rho _{01}\left( t\right) $ tends to $\rho
_{10}\left( 0\right) +\rho _{01}\left( 0\right) =p+q-2pq.$when $t$ tends to
infinity. Thus the stationary matrix of composite system after such
manipulation takes the form:%
\begin{equation}
\rho \left( \infty \right) =%
\begin{pmatrix}
pq &  &  &  \\ 
& 0 &  &  \\ 
&  & p=q-2pq &  \\ 
&  &  & \left( 1-p\right) \left( 1-q\right)%
\end{pmatrix}%
\text{.}  \label{v11}
\end{equation}%
The expression Eq. (\ref{v11}) implies that the state of first subsystem is $%
\rho _{1}=%
\begin{pmatrix}
pq &  \\ 
& 1-pq%
\end{pmatrix}%
$ and the state of second one is $\rho _{2}=%
\begin{pmatrix}
p+q-pq &  \\ 
& \left( 1-p\right) \left( 1-q\right)%
\end{pmatrix}%
$. Thus required task is entirely completed. It is easy to verify that if we
replace the above operator $R=\sqrt{\frac{a}{2}}f_{2}^{+}f_{1}$ by the
operator -$\sqrt{\frac{a}{2}}f_{2}f_{1}^{+}$ we come to the result that :$%
\rho _{1}=\rho \left( A\text{ }or\text{ }B\right) $ and $\rho _{2}=\rho
\left( A\text{ and }B\right) $ that is the states of the subsystems will be
swapped. Let us examine the another example that demonstrates how with the
help of quantum engineering tools one can copy different PP. Let us consider
two-qubit quantum system (with $H=0$ ) whose interaction with environment is
determined by two operators:$R_{1}=\sqrt{\frac{a}{2}}f_{2}^{+}N_{1}$ and $%
R_{2}=\sqrt{\frac{b}{2}}f_{1}^{+}N_{2}.$Then the appropriate Lindblad Eq. (%
\ref{v2}) for the diagonal elements of $\rho \left( t\right) $ in this case
takes the form:%
\begin{widetext}
\begin{equation}
\frac{d\rho _{N_{1}N_{2}}}{dt}=a\left( N_{2}N_{1}\rho _{N_{1}\overline{N_{2}}%
}-N_{1}\overline{N_{2}}\rho _{N_{1}N_{2}}\right) +b\left( N_{1}N_{2}\rho _{%
\overline{N_{1}}N_{2}}-\overline{N_{1}}N_{2}\rho _{N_{1}N_{2}}\right) \text{.%
}
\label{v12}
\end{equation}
\end{widetext}\bigskip\ or in component-wise form Eq. (\ref{v12}) reads as:%
\begin{eqnarray}
\frac{d\rho _{00}}{dt} &=&0  \notag \\
\frac{d\rho _{10}}{dt} &=&-a\rho _{10}\text{.}  \label{v13} \\
\frac{d\rho _{01}}{dt} &=&-b\rho _{01}  \notag \\
\frac{d\rho _{11}}{dt} &=&a\rho _{10}+b\rho _{01}  \notag
\end{eqnarray}%
The system of Eq. (\ref{v13}) implies that: $\rho _{00}\left( t\right) =\rho
_{00}\left( 0\right) =\left( 1-p\right) \left( 1-q\right) .$In addition both
elements $\rho _{10}\left( t\right) $ and $\rho _{01}\left( t\right) $ tend
to zero and $\rho _{11}\left( t\right) $ tends$\ $to$\ \ \rho _{11}\left(
\infty \right) =\rho _{10}\left( 0\right) +\rho _{01}\left( 0\right) +\rho
_{11}\left( 0\right) =p+q-pq.$ when t tends to infinity.Remind, that as
before we assume that tensor product matrix $\rho \left( A\right) \otimes
\rho \left( B\right) $ is the input state of the system.Thus we result in
that stationary DM of two-qubit system in question is%
\begin{equation}
\rho _{st}=%
\begin{pmatrix}
p+q-pq &  &  &  \\ 
& 0 &  &  \\ 
&  & 0 &  \\ 
&  &  & \left( 1-p\right) \left( 1-q\right)%
\end{pmatrix}%
\text{.}  \label{v14}
\end{equation}%
which implies that both subsystems are in the same state $\rho _{1}=\rho
_{2}=%
\begin{pmatrix}
p+q-pq &  \\ 
& \left( 1-p\right) \left( 1-q\right)%
\end{pmatrix}%
$. Thus the copying or doubling the LM of PP "$A$ or $B$" is realized. We
will not draw other examples because it is clear already that all logical
constructions of PL can be successfully simulated by the tools of quantum
engineering. It should be emphasized, as far as we involve only diagonal
elements of DM for implementation of logical operations, destructive effect
of decoherence will be far weaker that in standard quantum computation.We
remind that in the present paper the question is only about classical
computations with logical propositions in relevant quantum system.

\section{Application of probabilistic logic to the study of concrete
psychological problem}

In this part we examine the only illustrative example of application PL
methods to concrete problem, namely algebraic theory of human reflexive
choice. This theory was proposed many years ago by V.A. Lefebvre $\left[ 3%
\right] $ and to the present time it has received already several
experimental corroborations \cite{7b}. Let us briefly remind basic facts of
Lefebvre's theory.It operates with three substantial variables $x_{1},x_{2,}$
$x_{3}$ that determine a behavior of a person in a situation when he (or
she) has only two possible alternatives, namely positive or negative pole
for choice The variable $x_{1}$ describes environment tension \ which causes
a person to choose a positive pole .The variable $x_{2}$ reflects the
influence of past experience on a person to move towards the same pole .And
the variable $x_{3\text{ }}$ is the measure of personal intention to choose
a positive pole. According to the Lefebvre theory the output variable X
which determines the behavior of a person ( that is probability of his or
her choosing the positive pole) may be found from the following algebraic
equation%
\begin{equation}
X=x_{1}+x_{3}\left( 1-x_{1}\right) \left( 1-x_{2}\right) .  \label{v15}
\end{equation}%
The Eq. (\ref{v15}) implies that if for example all variables $x_{i\text{ \ }%
}$take with equal probabilities values from interval $\left[ 0,1\right] $,
the probability X for a person to choose positive pole is equal to $\frac{1}{%
2}+\frac{1}{2}\times \frac{1}{2}\times \frac{1}{2}=\frac{5}{8}=0,625.$
Various experiments carried out with many persons from different social and
age groups when the situation of bipolar choice took place confirm in whole
this result of Lefebvre theory \cite{6b}. Note that in the case when $%
x_{1},x_{2},x_{3\text{ }}$are usual Boolean variables the output variable X
can be represented as Boolean function of double implication%
\begin{equation}
X=(\left( x_{3}\Longrightarrow x_{2}\right) \Longrightarrow x_{1}).
\label{v16}
\end{equation}%
By the way this fact implies that proposition $\left( \left(
x_{1}\Longrightarrow x_{3}\right) \Longrightarrow x_{2}\right) $ and other
propositions with the same logical structure will result in (when variables
are distributed uniformly in interval $\left[ 0,1\right] $) the final result
as in the case of the Lefebvre theory. In present paper we propose certain
generalization of the above theory on the case when the situation of choice
leaves to a person three possible alternatives that is besides positive and
negative poles a person may for example to choose middle way or refrain from
choice making at all.If we assume that the logical structure of the basic
test equation Eq. (\ref{v16}) is conserved then we must only \ replace the
Eq. (\ref{v15}) by a similar one\ in which the double implication will be
expressed in terms of three valued PL. Using twice the relation Eq. (\ref{v6}%
) for LM of implication in three-valued PL and omitting the simple algebra
we come to the required result.The probability for a person to choose
positive pole in situation with three possible alternatives is%
\begin{equation}
X=p_{1}+r_{1}q_{3}-p_{1}q_{3}r_{1},  \label{v17}
\end{equation}%
and the probability to choose negative pole - $Y$ \ in such situation is%
\begin{equation}
\ Y=p_{3}\left( q_{1}+r_{3}-q_{1}r_{3}\right) .  \label{v18}
\end{equation}%
The remainder probability $Z$ to make a middle way choice can be determined
by obvious relation: $Z=1-X-Y$. In relations Eq. (\ref{v17}) and (\ref{v18})
we used the following notation: probabilistic proposition $x_{1}$ has LM -$%
\rho \left( x_{1}\right) =%
\begin{pmatrix}
p_{1} &  &  \\ 
& p_{2} &  \\ 
&  & p_{3}%
\end{pmatrix}%
,$ PP-$x_{2}$ has LM-$\rho \left( x_{2}\right) =%
\begin{pmatrix}
q_{1} &  &  \\ 
& q_{2} &  \\ 
&  & q_{3}%
\end{pmatrix}%
$and respectively PP has LM of the form $\rho \left( x_{3}\right) =%
\begin{pmatrix}
r_{1} &  &  \\ 
& r_{2} &  \\ 
&  & r_{3}%
\end{pmatrix}%
$. In the case when all probabilities $p_{i},q_{i},r_{i}$ are equal to each
other ( and equal to$\frac{1}{3}$) the required output probabilities are:
for positive choice $X=\frac{11}{27}$ and for negative choice $Y=\frac{5}{27}%
.$Respectively the probability of the middle way choice \ is equal to $\frac{%
11}{27}$. We believe that expressions Eq. (\ref{v17}) and (\ref{v18}) can be
verified directly by relevant psychological experiments and after such
testing it will became more clearly how universal is the implicative logical
structure of the Lefebvre theory .

Let us summing up the results of our investigation.Based on ideas of QTOS we
propose the consistent approach which allows one to operate with plausible
propositions according to the rules that generalisize the laws of ordinary
Boolean calculus.This approach can be extended also on the case of
many-valued logic and gives one the possibility to consider in the framework
of unified method operations with propositions with different number of
outcomes.We hope that approach proposed in view of its generality and
flexibility will find applications for the study of various concrete
problems both in physics and in "soft" sciences.

In conclusion it should be noted that main driving motive for the author to
write this parer was his belief that well-known R. Landauer's thesis
"Information is physical" can be applied not only to information but to
logic as well.

I would like to aknowledge L.A.\ Pastur for useful discussions of the
results of this paper.

\end{document}